\documentclass[conference]{IEEEtran}
\IEEEoverridecommandlockouts

\usepackage{cite}
\usepackage{amsmath,amssymb,amsfonts}
\usepackage{algorithmic}
\usepackage{graphicx}
\usepackage{textcomp}
\usepackage{xcolor}

\usepackage{hyperref}
\hypersetup{
    colorlinks=true,
    urlcolor=[RGB]{25,168,255},
    citecolor=[RGB]{25,168,255},
    linkbordercolor = {white}
}
\usepackage{cite}

\title{HPC Extensions to the OpenKIM Processing Pipeline\\
  \thanks{This work is supported by the National Science Foundation under awards DMR-1834251 and DMR-1834332}
}

\makeatletter
\newcommand{\linebreakand}{%
  \end{@IEEEauthorhalign}
  \hfill\mbox{}\par
  \mbox{}\hfill\begin{@IEEEauthorhalign}
}
\makeatother

\author{\IEEEauthorblockN{Daniel S.\ Karls}
\IEEEauthorblockA{
Department of Aerospace Engineering\\
and Mechanics\\
University of Minnesota\\
Minneapolis, MN 55455 \\
ORCID: 0000-0002-4069-396X}
\and
\IEEEauthorblockN{Steven M.\ Clark}
\IEEEauthorblockA{San Diego Supercomputer Center\\
University of California, San Diego\\
La Jolla, CA 92093\\
ORCID: 0000-0001-9345-4682}
\and
\IEEEauthorblockN{Brendon A.\ Waters}
\IEEEauthorblockA{
Department of Aerospace Engineering\\
and Mechanics\\
University of Minnesota\\
Minneapolis, MN 55455 \\
ORCID: 0000-0001-7324-3909}
\linebreakand
\IEEEauthorblockN{Ryan S.\ Elliott}
\IEEEauthorblockA{
Department of Aerospace Engineering\\
and Mechanics\\
University of Minnesota\\
Minneapolis, MN 55455 \\
ORCID: 0000-0003-4988-8306}
\and
\IEEEauthorblockN{Ellad B.\ Tadmor}
\IEEEauthorblockA{
Department of Aerospace Engineering\\
and Mechanics\\
University of Minnesota\\
Minneapolis, MN 55455 \\
ORCID: 0000-0003-3311-6299}
}
\begin{document}

\maketitle

\begin{abstract}
The Open Knowledgebase of Interatomic Models (OpenKIM) is an NSF Science Gateway that archives fully functional computer implementations of interatomic models (potentials and force fields) and simulation codes that use them to compute material properties.
Interatomic models are coupled with compatible simulation codes and executed in a fully automated manner by the OpenKIM processing pipeline, a cloud-based computation platform.
The pipeline as previously introduced in the literature was insufficient to support the large-scale computations that have become necessary within the materials science community.
Accordingly, we present extensions made to the pipeline that allow it to utilize High-Performance Computing (HPC) resources in an efficient and performant fashion.
\end{abstract}


\section{Introduction}
In recent years, the materials science community has increasingly focused on numerical modeling approaches to predict the properties of novel materials.
While fully determining the electronic states of a material via quantum mechanical approaches like density functional theory (DFT) remains prohibitively expensive to calculate for anything but the smallest systems, researchers have turned to atomistic methods where the interactions are computed using approximate interatomic models (IMs) (empirical interatomic potentials and force fields) for the increasingly large sections of materials typically modeled in contemporary materials science research. This trend has accelerated even further with recent developments in machine learning that have led to the development of more accurate IMs.
Many IMs have been developed for a plethora of materials over the last few decades that vary considerably in their complexity, computational efficiency, and transferability across different material properties.
Furthermore, the computer implementation of these models is often not published with the corresponding research.
This can make it difficult for subsequent researchers interested in similar systems to determine whether a model that fits their needs has already been developed, or which of several existing IMs most accurately predicts a particular physical property relevant to their investigation.

The Open Knowledgebase of Interatomic Models (OpenKIM, KIM)~\cite{tadmor:elliott:2011,tadmor:elliott:2013} is an NSF-funded Science Gateway or cyberinfrastructure~\cite{SGCI,XSEDE} founded in 2009 to address the need for a central repository of IMs to help researchers find applicable models and make informed comparisons between them.
OpenKIM archives fully functional computer implementations of IMs with the relevant parameters, maintained in a public database hosted at \url{https://openkim.org}.
Archived alongside the models are a host of property computations, i.e.\ simulation codes that are coupled with them to compute material property predictions, called \emph{Tests} in OpenKIM parlance.
This coupling is made possible by the KIM Application Programming Interface (KIM API)~\cite{kim-api}, which provides a standard mechanism for models and simulations to exchange information.
When a new IM is uploaded to OpenKIM, it is automatically paired and executed with all compatible property computations and the resulting material property predictions are stored in the database; similarly, when a new property computation is uploaded, it is run against all compatible IMs.
In general, property computations may query the OpenKIM database for the results of other computations, which imposes a partial ordering on the sequence in which they must be run.
Carrying out these computations is, therefore, a non-trivial endeavor requiring careful implementation in addition to computational resources.
Once completed, their results allow for direct comparisons to be made between the predictions of comparable models (e.g., which IMs best predict the geometric parameters of the most common bulk phases of silicon?).

This process of taking an IM submitted to openkim.org, automatically determining which property computations are compatible with it to form jobs, resolving the dependencies between property computations, executing the jobs, and presenting their results is handled by the OpenKIM \emph{processing pipeline}~\cite{karls-openkim-pipeline-jcp-2020}.
The pipeline is composed of several high-level components connected by asynchronous Celery~\cite{celery} task queues running RabbitMQ~\cite{rabbitmq} as a message broker.
In its original design, there were three main components: (1) the \emph{Gateway}, responsible for handling communication between openkim.org and the other components; (2) the \emph{Director}, which creates jobs for compatible model-simulation code pairs and determines the order in which they are run, respecting their dependencies; and (3) an array of \emph{Workers} responsible for carrying out the actual calculations.
These Workers were run on commodity hardware, and were thus limited to running small jobs, each containing a small number of simultaneous processes.
To enable the OpenKIM project to grow to meet the needs of the materials science research community, there is a need to be able to submit jobs to third-party High-Performance Computing (HPC) clusters to take advantage of large-scale computational resources that already exist.
This not only allows for larger individual jobs, but also for a greater quantity of jobs to be run simultaneously.
However, this capability comes at the cost of increased complexity, the details of which will be explored in the remainder of this article.

\section{Design and Implementation}

\subsection{Executing jobs on HPC resources}
The fundamental unit of the OpenKIM pipeline is the \emph{KIM Job}.
A KIM Job is a single coupling of an IM and a property computation (Test) used to compute a specific physical property.
Each job needs to run in a controlled, reproducible software environment that includes any external packages required by the simulation.
This environment must also be easily portable and require minimal permissions to run on HPC clusters, which typically only allow user-level access.
The pipeline addresses these needs by executing KIM Jobs inside of Singularity~\cite{singularity} containers.
In addition to ensuring that jobs run with all required software, the isolation provided by containers also makes it possible to prevent multiple jobs running simultaneously on the same compute node from interfering with one another.

With a viable way of running jobs on HPC clusters established, a second challenge is to devise a method of taking KIM Jobs, which are generated within the pipeline, and submitting them to the scheduler of a given cluster.
To this end, the pipeline utilizes the Tapis~\cite{Tapis} platform operated by the Texas Advanced Computing Center (TACC).
Tapis is a web-based API framework intended to allow researchers to flexibly and securely orchestrate computational workflows across heterogeneous HPC resources.
Users begin by registering any desired storage and execution systems to which they have access, and proceed to define \emph{Tapis Apps} that work atop them.
Each Tapis App is required to have its own well-defined set of parameters and input files that it needs to run.
A \emph{Tapis Job} is then created by invoking a Tapis App with explicit values for its parameters and the paths of specific input files on a registered storage system.
Once a Tapis Job is submitted, it is possible to subscribe to event notifications as it progresses through a number of different states in its life cycle, allowing it to be monitored without repeatedly polling HPC resources directly.
Finally, once a Tapis Job has completed, all files present in its directory on the execution system, including any new files it produced while running, can be copied back to a specified storage system.

As described above, Tapis promises to provide a unified, abstract interface for the pipeline to interact with HPC resources.
However, before KIM Jobs can be submitted to clusters as Tapis Jobs, two changes must be made to the workflow of the pipeline.
First, each KIM Job must be assigned a number of CPU cores and an execution time to request.
This information is defined on a per-Test basis: when a new property computation is submitted to OpenKIM, the submitter indicates the core count and time that each KIM Job associated with it should request.
The Workers of the original pipeline architecture must also be replaced by a new component that serves as the interface between Tapis and the rest of the pipeline.
For this purpose, the \emph{Dispatcher} component is introduced.
Rather than being fetched and executed by Workers, new KIM Jobs spawned by the Director are received by the Dispatcher, which packages\footnote{The packaging of KIM Jobs includes performing any queries for other job results that may be required by the property computation and storing the information in an input file accompanying the job. This is done because external network access is generally not allowed from HPC clusters.} them into a format expected by the Tapis App used by the pipeline and submits them as Tapis Jobs.

\subsection{Scalable job submission}
Given the large number of KIM Jobs that pass through the pipeline, submitting each as an individual Tapis Job would be impractical for two reasons.
First, the high volume of jobs per unit time would place undue stress upon both the Tapis service and the cluster schedulers.
Second, the compute nodes on modern execution systems can contain a relatively large number of cores and node sharing may not be allowed by the scheduler, i.e.\ only a single job may be running on a given compute node at any given time.
Jobs in the pipeline range significantly in the number of cores they request, and reserving an entire compute node only to use a small fraction of its cores to run a single KIM Job is highly inefficient.

\begin{figure*}[t]
    \centering
    \includegraphics[width=0.6\textwidth]{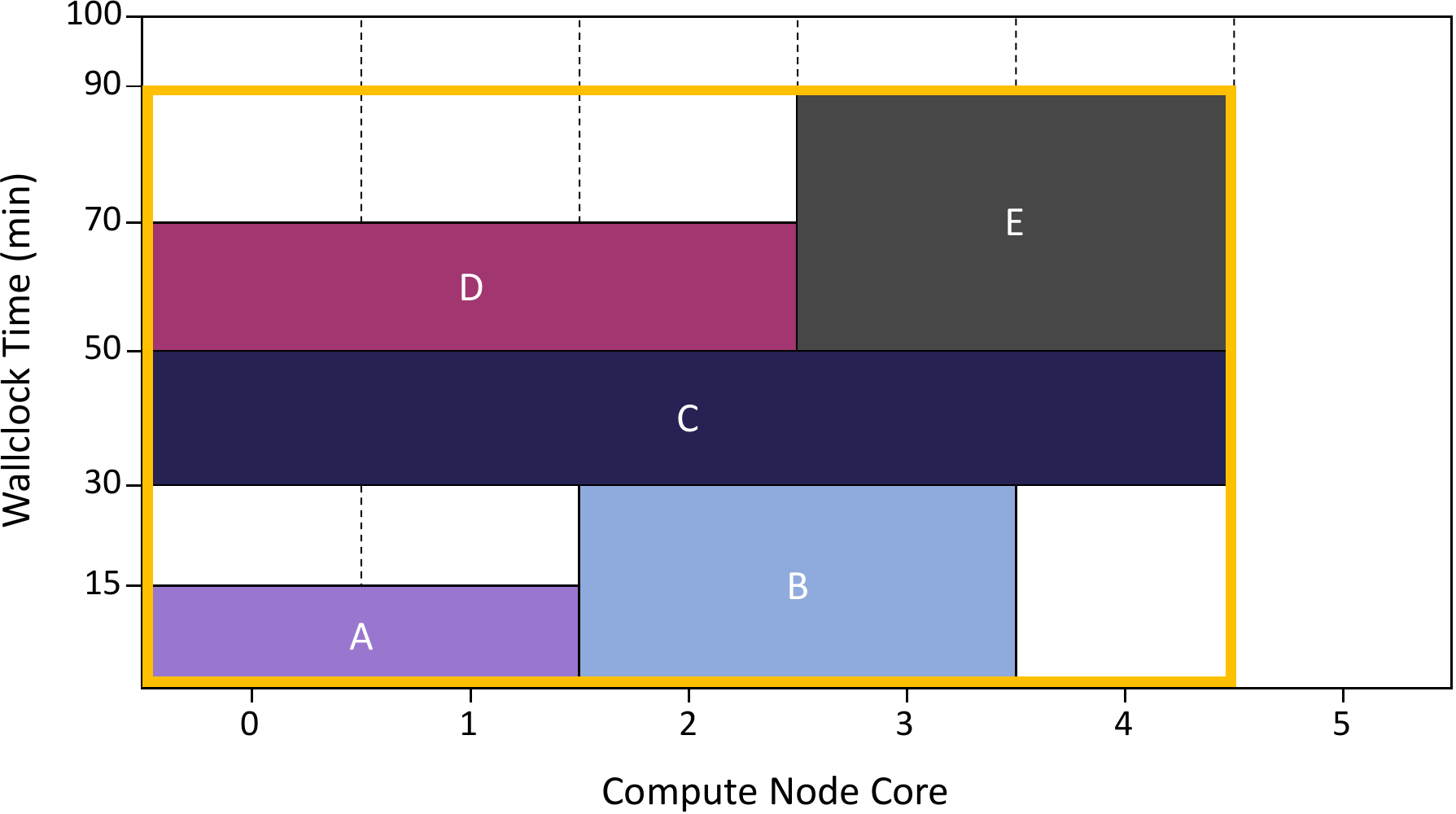}
    \caption{
      An example of bundling five KIM Jobs using the MAXRECTS-BL algorithm for an execution site whose compute nodes feature six cores and for which jobs are allowed at most 100 minutes of execution.
      The yellow rectangle reflects the total number of cores and execution time that will be requested for the bundle.
      White spaces within the bundle indicate wasted compute time; this is, in part, due to the fact that jobs are placed in the bin online in the order in which they were received (A$\rightarrow$B$\rightarrow$C$\rightarrow$D$\rightarrow$E) and not rearranged during packing.
    }
    \label{fig:bundling}
\end{figure*}

To remedy this problem, in practice the Dispatcher groups KIM Jobs into \emph{bundles} that are each submitted as a separate Tapis Job.
Associated with each HPC cluster accessible to the pipeline are a set of scheduler queues or \emph{partitions} that can be used to run jobs.
These partitions vary in terms of the specific type of compute node they run jobs on, which has an explicit number of CPU cores, as well as the maximum allowable wallclock time for jobs submitted to them.
We refer to each HPC cluster-scheduler partition pair as an \emph{execution site}.
The Dispatcher uses an internal SQLite~\cite{sqlite} database to store the aforementioned information for each execution site.
Accompanying each execution site in the database is a batch queue to be filled with KIM Jobs.
When the Dispatcher first receives a new KIM Job, it queries its database for all available execution sites that can accommodate the number of cores and run time requested by the job.
As a primitive load balancing mechanism, one of these execution sites is selected at random (with equal probability) and the job is bound to the relevant queue.
The Dispatcher then checks to see whether the jobs currently bound to that queue suffice to form a bundle, which we discuss next.

\begin{figure*}[t]
    \centering
    \includegraphics[width=\textwidth]{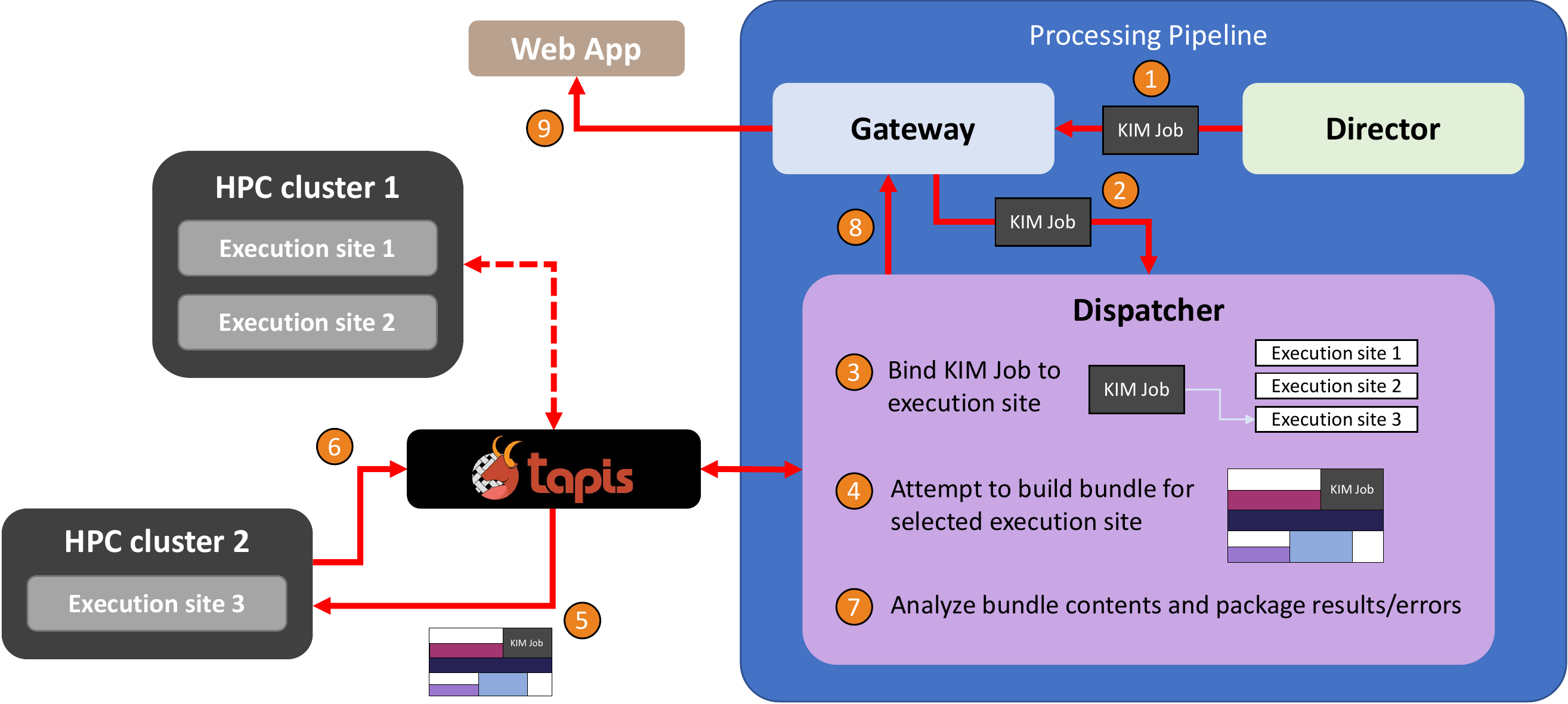}
    \caption{
      Overview of the workflow of the HPC-enabled pipeline: (1) a new KIM Job is created by the Director and put on a job queue on the Gateway; (2) the Dispatcher retrieves the job and (3) binds it to an execution site; (4) an attempt is made to bundle the new KIM Job with all other KIM Jobs currently bound to its execution site since the last time a bundle was created; (5) assuming a bundle of sufficient size can be made, it is submitted as a single Tapis Job and copied onto its execution site, where it is submitted to the cluster scheduler and run; (6) once execution of all KIM Jobs in the bundle has completed, the contents of the bundle's directories are copied back to the Dispatcher by Tapis; (7) the Dispatcher inspects the contents of the bundle and packages results or errors and (8) they are copied back to the Gateway before (9) being sent to the OpenKIM web app for display on openkim.org.  We refer the reader to~\cite{karls-openkim-pipeline-jcp-2020} for further details of the Gateway, Director, and web app.
    }
    \label{fig:hpc_pipeline_overview}
\end{figure*}

Because each execution site has a specific core count and wallclock time associated with it, it can be thought of as defining an operable region or \emph{bin} in a two-dimensional resource space whose (discrete) horizontal axis corresponds to CPU cores and whose vertical axis corresponds to execution time.
This rectangular region can be filled with KIM Jobs, each of which similarly defines a rectangle in this space.
This application is a simple example of the more general ``knapsack problem'' in combinatorial optimization~\cite{knapsack_problems}.\footnote{In the general knapsack problem, multiple bins are packed simultaneously from the same set of objects, with the objective being to produce a packing requiring the smallest number of bins.}
Even for a single two-dimensional bin, and without allowing rectangles to rotate during packing since our resource space axes are not equivalent, various objective functions and methods can be used.
We choose the Maximal Rectangles-Bottom Left (MAXRECTS-BL) algorithm described in~\cite{MAXRECTS-BL}, as implemented in the \texttt{rectpack}~\cite{rectpack} python package.
When a given rectangle is being packed into the bin, a set of candidate placements is first formed for which the rectangle will not overlap with any existing rectangles in the bin and for which its upper edge has the lowest possible value along the vertical axis (execution time).
From these candidates, the placement for which the left edge of the rectangle possesses the smallest value along the horizontal axis is selected.
The reasoning for first prioritizing that rectangles be placed toward the bottom of the bin in our application is that it maximizes both throughput and efficiency.
All KIM Jobs contained in the same bundle are independent of one another, and so can be executed in parallel to minimize the overall time to solution.
In light of the aforementioned lack of node sharing on many HPC systems, it is also most efficient to occupy as many cores of a compute node simultaneously as possible for the duration of the bundle's execution.
We then prioritize having rectangles placed as far left in the bin as possible to ensure that, if the bundle is being submitted to an execution site that does feature node sharing, we request only as many cores as will actually be used.

The bundling procedure for a given execution site outlined above is carried out in an online manner: KIM Jobs bound to the relevant execution site are iterated over one-by-one in the order in which they were received by the Dispatcher.
If a KIM Job's rectangle fits in the bin for the execution site, it remains fixed in the position in which it was placed for the duration of the packing process.
As soon as a sufficient number of KIM Jobs have been packed into the bin, the iteration ceases, and a job bundle is formed and submitted to the execution site as a single Tapis Job.
The final number of cores and run time requested for the job correspond to the smallest rectangular region that fits inside of the bin but still encompasses all of the KIM Job rectangles.
This is illustrated in the toy example of our chosen method of bundling shown in Fig.~\ref{fig:bundling}.
If there were not enough KIM Jobs that were able to be packed into the bin, the bundling process is aborted, to be repeated the next time a new KIM Job is bound to the relevant execution site.\footnote{If no new KIM Jobs have been received by the Dispatcher for some time, those currently bound to an execution site that lacks a sufficient number of jobs to create a bundle may be left idle.  To avoid this, bundles are created for each execution site at regular time intervals regardless of how many KIM Jobs are currently bound to them.}

Once on the cluster, Tapis submits the bundle to the scheduler as a single overarching job that spawns one subprocess, or \emph{step}, for each constituent KIM Job.
These steps are executed in accordance with the position in which the KIM Jobs were packed within the bundle: a KIM Job cannot begin executing until all KIM Jobs that fall beneath it have finished.
This on-node execution scheduling is implemented using the GNU \texttt{make} utility~\cite{GNUmake} by leveraging its mechanism for dependencies between targets.
For the example bundle shown in Fig.~\ref{fig:bundling}, each of the five KIM Jobs defines one target that contains a command to execute the job in its corresponding Singularity container.
The targets for KIM Jobs A and B have no dependencies, while the target for job C lists both A and B as dependencies; the targets for jobs D and E both list job C as a dependency.
In this case, the overarching bundle job begins by executing the steps for jobs A and B in parallel.
The step for job C waits until A and B were finished running before being started.
Finally, once job C is complete, jobs D and E begin executing in parallel.
Although cores on the compute node are not explicitly reserved for specific steps at any point, this method of execution ensures that each step receives, on average, the number of cores it requested for the duration of the wallclock time it requested.
Once all steps are done executing, the contents of the bundle's directory are copied back to the Dispatcher, which analyzes scheduler accounting information for the bundle and the subdirectory for each KIM Job.
The results of KIM Jobs that managed to complete are packaged into either results or errors and sent back to the Gateway, which inserts them in the primary database and forwards them to the OpenKIM web app for display on openkim.org.
A schematic of the entire process is shown in Figure~\ref{fig:hpc_pipeline_overview}.

\section{Robustness considerations}
Implicit in the workflow described above are several technical complications that warrant discussion.
First, the execution time requested by a KIM Job may prove insufficient, resulting in a timeout on the cluster.
In such an event, different schedulers vary in their precise behavior depending on how they are configured, but common among them is that a job is sent a termination signal and is given a finite duration of time to tear itself down; if the job is still active after this period has passed, it is forcibly killed.
In deciding the time allocated to each step inside of a bundle, a small buffer of at least this magnitude must be added to the time originally requested by the corresponding KIM Job in anticipation of a possible timeout.
It is also critical that the Dispatcher be able to properly detect timeouts of steps within bundles when it analyzes the contents of the bundle directory returned by Tapis.
This is accomplished by examining the scheduler accounting information for a bundle, which is printed to a file after all steps have finished executing and the scheduler has had time to poll their status.
Upon detecting a timeout, the Dispatcher ensures that when the relevant KIM Job is next submitted, it will request twice as much wallclock time as it previously did.
However, increasing the requested time for a job may mean that the execution site to which it was originally bound can no longer accommodate it.
In this case, it is rebound to a new execution site, chosen randomly out of those remaining that are compatible with its requested resources.
This repeats until a job is either able to finish running or its requested resources exceed those of any available execution sites, at which point a designated error is packaged for it that is returned to the Gateway.

Aside from timeouts, KIM Jobs may fail to execute because of hardware-related issues.
These issues can range in scope from being localized to one or more compute nodes to affecting an entire cluster, e.g.\ a power failure or widespread network outage.
Further, all systems, including Tapis itself, must undergo planned maintenance periodically.
This may occur before an attempt is made to submit a bundle for execution or while a bundle is already executing.
To account for such failure modes, an indirect approach is taken.
When the Dispatcher analyzes the contents of a bundle returned by Tapis, it excludes the possibility of a localized hardware malfunction by verifying the existence of certain sentinel files that are always created for each job at its conclusion.
To determine whether a global issue has occurred, the Dispatcher periodically inspects all active Tapis Jobs in its database.
Every time a notification is received for a Tapis Job, it is recorded along with a timestamp.
If the time elapsed since a notification was last received for a job significantly exceeds the total wallclock time it requested, it is cancelled and resubmitted as part of a new bundle.
In the event that an HPC cluster is experiencing frequent or prolonged outages, the pipeline also features the ability to deactivate execution sites in real time, which automatically rebinds their jobs to compatible sites still active.
Once their availability has stabilized, they can be reactivated and accept jobs once more.

\section{Conclusion and future work}
In this paper, we discuss an extension to the OpenKIM processing pipeline described in~\cite{karls-openkim-pipeline-jcp-2020}, whose purpose is to couple IMs with compatible property computations and execute the corresponding jobs, respecting any dependencies that may exist between them.
In its original design, the pipeline was limited to running small jobs consisting of at most several simultaneous processes each.
Prompted by the need to support execution of large-scale jobs, we have extended the pipeline to run jobs on multiple HPC resources with the ability to easily extend to more.
This is achieved by replacing the Workers present in its original architecture with a new component, the Dispatcher, that forms bundles of jobs for each HPC resource and submits them using the Tapis platform.
Upon reaching the cluster, the jobs are executed in a portable, isolated fashion using containers in a sequence derived from their relative position within the encompassing bundle.
Throughout the process of executing jobs, timeouts and other modes of failure are handled gracefully by the Dispatcher, including dynamically rebinding jobs to different HPC resources as necessary.

With regard to future work, one area where significant gains in efficiency could be made is in the resources requested for each KIM Job.
Currently, the same wallclock time is requested for the initial submission of every KIM Job that uses a given simulation code.
However, the actual execution time of every completed job is recorded by the Dispatcher, and statistical analysis of this information could conceivably allow for more accurate estimates for the expected run time of new KIM Jobs based on their constituent property computation or IM.
Such analysis might also reveal a more sophisticated mechanism of load balancing between HPC resources.
Because the different HPC resources used to run jobs are heterogeneous, a given job may take longer to execute on one system than another.
Accordingly, the current method of equally distributed load balancing between execution sites may be suboptimal, although other considerations such as the time jobs spend queued before beginning to run would also need to be considered.
Finally, efficiency could be improved by using an offline packing algorithm when creating bundles to more exhaustively search the space of possible packings for a given execution site bin and set of KIM Jobs.

\section{Acknowledgments}
The authors wish to acknowledge support from the Science Gateways Community Institute (SGCI)~\cite{SGCI}.
This work also used the Extreme Science and Engineering Discovery Environment (XSEDE)~\cite{XSEDE}, which is supported by National Science Foundation grant number ACI-1548562.
Specifically, this work used the Stampede2 cluster at the Texas Advanced Computing Center (TACC) and Jetstream\cite{jetstream} services hosted at Indiana University and TACC through allocation TG-MAT200008.
Finally, we are grateful to XSEDE for enabling S.~M.\ Clark to work with us through the Extended Collaborative Support Service (ECSS)~\cite{ECSS} program. This work is supported by the National Science Foundation under awards DMR-1834251 and DMR-1834332.

\bibliography{IEEEabrv,hpc_openkim_pipeline}

\end{document}